\documentclass{article} 
\usepackage{iclr2020_conference,times}

\usepackage{amsmath,amsfonts,bm}

\def\eqref#1{equation~\ref{#1}}

\def\1{\bm{1}}

\DeclareMathAlphabet{\mathsfit}{\encodingdefault}{\sfdefault}{m}{sl}
\SetMathAlphabet{\mathsfit}{bold}{\encodingdefault}{\sfdefault}{bx}{n}

\usepackage{hyperref}
\usepackage{url}
\usepackage{graphicx}

\title{Advancing Renewable Electricity \\ Consumption With Reinforcement Learning}

\author{Filip Tolovski \thanks{ Additional email address: tolovskifilip@gmail.com} \\
Fraunhofer Institute for Telecommunications, HHI \\
10587 Berlin, Germany  \\
\texttt{filip.tolovski@hhi.fraunhofer.de} \\
}

\iclrfinalcopy 
\begin{document}

\maketitle

\begin{abstract}As the share of renewable energy sources in the present electric energy mix rises, their intermittence proves to be the biggest challenge to carbon free electricity generation. To address this challenge, we propose an electricity pricing agent, which sends price signals to the customers and contributes to shifting the customer demand to periods of high renewable energy generation. We propose an implementation of a pricing agent with a reinforcement learning approach where the environment is represented by the customers, the electricity generation utilities and the weather conditions. 

\end{abstract}

\section{Introduction}

The intermittence of renewable energy sources in the present electric energy generation systems poses an issue to  electric utilities and electricity markets. These are built on the notion that supply follows demand \citep{NREL_overgeneration}. Their integration in the present energy markets, due to the lack of scalable storage solutions \citep{ValueStorage}, can be most efficiently addressed by demand response \citep{osti_1097911}.  
Demand response defines changes in the electric usage by consumers from their normal preferences caused by changes in  electricity  prices or by other incentives \citep{international2003power}. 
With demand response through real time pricing \citep{international2003power}, utilities can shift the customer load demand to periods of excess renewable energy generation. 
These peaks of generated energy are caused by high renewable energy generation, while the load demand is low. The shifting of customer load demand can result with reducing the greenhouse gasses emissions by natural gas and coal electricity generation plants, used at times of peak demand or for back up generation \citep{osti_1097911}.

\section{Related work}
\label{gen_inst}

Previous research on using reinforcement learning for dynamic pricing in hierarchical markets has shown promising results, both in balancing of the supply and demand, as well as in cutting the costs for customers and electric utilities alike \citep{DP_DR_RL_approach}. If a reinforcement learning model for customer load scheduling is also included, there is a cost reduction on the both sides when compared to \textit{myopic optimization} \citep{DP_Scheduling_RL}. However, there is no application of these methods in a physical environment. This is due to the lack of a simulation environment which can provide a reliable estimate of the safety and cost of the proposed method. To this end, we propose a simulation environment and adding weather data as an input to our proposed agent. 

\section{Reinforcement Learning Approach}
\label{headings}

\begin{figure}[t]
    \begin{center}
        \includegraphics[width=14 cm]{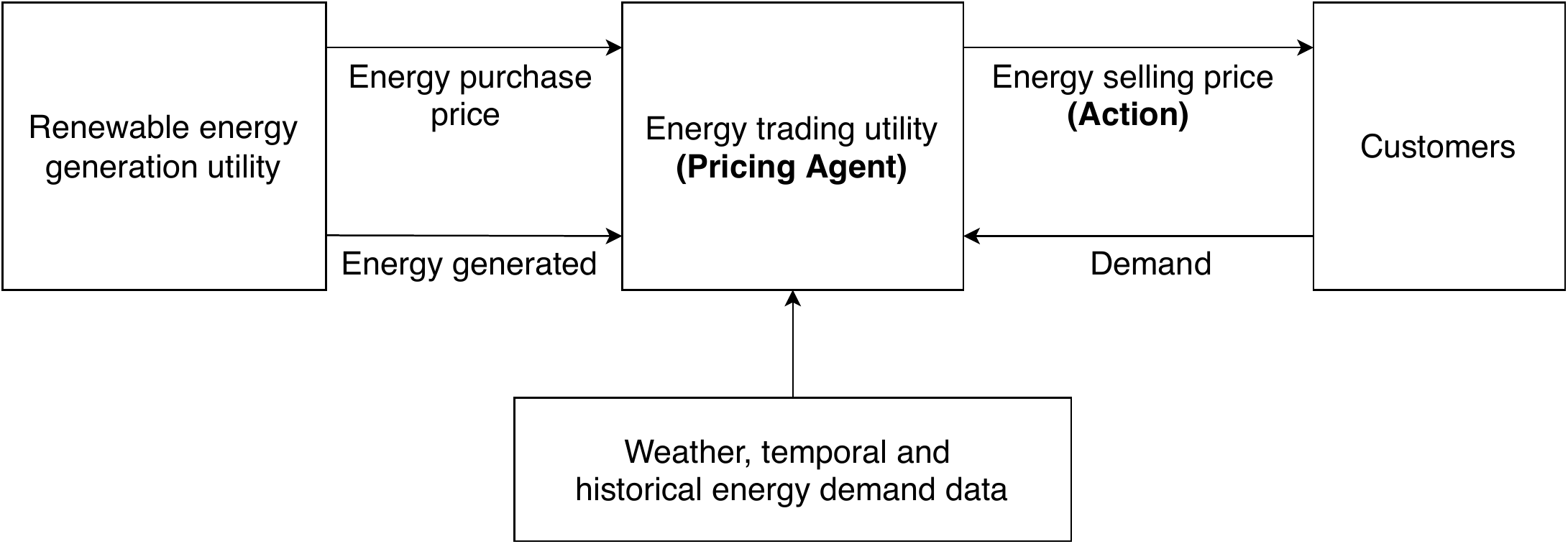}
        \caption{Diagram of the reinforcement learning setting. The pricing agent provides the new electricity selling price to the customers based on the received input data.}
        \label{fig:Diagram}

    \end{center}
\end{figure}

To model this problem as an reinforcement learning problem \citep{sutton2018reinforcement}, we choose the electric utility to represent the pricing agent. 
For the state \textit{s}\textsubscript{t}, we choose to be represented by the momentary and future renewable electricity supply to the electric utility, the momentary customer load demand together with the momentary and future features of the real world which are used for electricity demand forecasting \citep{DRMLtwostage, DLenergyprediction, demandforecasting}. At each timestep \textit{t}, the electric utility as the agent selects an action \textit{a}\textsubscript{t}, which is represented by the momentary and future electricity prices. This action is then transmitted back to the customers, which as a part of the environment responds with a load demand. This load demand is then used to calculate the reward \textit{r}\textsubscript{t}. Now, the pricing agent  is in another state \textit{s}\textsubscript{t+1} and this whole process is repeated for the duration of the simulation period, which is previously chosen. The environment and the agent are shown in Fig. \ref{fig:Diagram}.

Since it is very often that the renewable electricity sources are not in the vicinity of customers, we set the future supply to be given as an input.
The momentary demand, momentary and future renewable energy, its price, weather data and the temporal data are of dimensions $P+1$, while the energy selling price, i.e. the action, is of dimension 1. This way, the pricing agent formulates the price having as input the expectation for the $P$ future states. The size of the timestep and $P$ are treated as hyperparameters of the learning problem.

\subsection{Reward function}
The two objectives of the proposed pricing agent is to decrease the difference between the supply of renewable energy and demand and to keep the energy utility profitable. We propose the global reward function to be a linear combination of two sub-rewards as in the multi-objective analysis of return, shown in Eq. (\ref{eq:sum_reward}), proposed by \citet{RLchallenges}. 

\begin{equation}
    \label{eq:sum_reward}
    r\left ( s,a \right ) = \sum_{j=1}^{2}\alpha_j r_j\left ( s,a \right )
\end{equation}

where:

\begin{equation}
    \label{eq:r1}
    r_1\left ( s,a \right ) = \left (\ Price_{sold}\left ( s,a \right ) - Price_{purchased} \right )
\end{equation}

and 

\begin{equation}
    \label{eq:r2}
    r_2\left ( s,a \right ) = - \left ( Energy_{renewable} - E_{demand}\left (  s,a \right ) \right )^{2}
\end{equation}

The coefficients $\alpha_j$ in (\ref{eq:sum_reward}) represent hyperparameters of the learning problem and are initially set to 1. The sub-reward function shown in (\ref{eq:r1}) calculates the profit for the pricing agent as the difference between the purchase price and the selling price. The sub-reward function shown in (\ref{eq:r2}) calculates the square of the difference between the renewable energy available and the demand. The negative sign and the square form of the sub-reward function (\ref{eq:r2}) show the objective of reducing both, the positive and the negative difference between the renewable supply and demand.

\section{Implementation}

\subsection{Simulation environment}
Since real time testing of the proposed model is costly and there isn't available dataset for batch off-line training, we propose building a new simulation environment. In the proposed environment, customers are represented by a number of previously trained demand response agents. As a training ground for these agents, we propose the CityLearn \citep{citylearn}, an OpenAI Gym environment \citep{openaigym}. 
CityLearn enables training and evaluation of autonomous and collaborative reinforcement learning models for demand response in resident buildings. 
In order to be able to model buildings which are not fitted with energy storage capabilities, we remove the storage capabilities of some of the buildings in CityLearn. 
The cost function for the customer agents is chosen to minimize both the peak energy demand and the cumulative energy cost. To improve the generalization of the pricing agent with emerging smart cities and neighbourhoods, we propose training both independent and cooperative customer agents. The independent agent is not aware for the actions of other agents, while the cooperative agent value its actions in conjunction with the actions of other agents \citep{jal}. 
After evaluating the performance of the customer agents, a number of different simulation environments are built by combining trained customer agents. The distribution of customer agents in an environment is set as such to ensure that the pool of simulation environments properly models the momentary and future price responsiveness of the physical environment. 

\subsection{Training}
The training of the pricing agent is done in two phases. In the first phase, the pricing agent is trained using model-agnostic meta-learning \citep{maml} across all simulation environments. The second phase, where the pricing agent is trained and ran in the physical environment, is started after reaching a certain performance threshold in the simulation environment.
The training in the first phase should increase the sample efficiency of the pricing agent in the second phase and should reduce the costs and the risks of training in the physical environment. 
Training on multiple environments with different distributions of customer agents should also increase the robustness of the pricing agent in the second phase \citep{RLchallenges}. 

\subsection{Safety and Explainability}
To ensure the safety of the pricing agent operating in a physical environment, we propose using a constraints on the price it signals to the customers. The value of the constraints should be set by the electricity utility, according to their pricing policy and to the market regulations. In order to evaluate the safety of an algorithm in respect to the constraints, we propose using summary of all violations to the constraints, such as in \citet{safety}. Regarding evaluation of the impact these constraints have on the performance of the pricing agent, we propose learning a policy as a function of the constraint level as in \citet{budget_MDP_1,budget_MDP_2}. This should provide information about the trade-offs between the constraint level and the expected return to the human operators of the pricing agent \citep{RLchallenges}.
In order to further improve the explainability of the pricing agent, we propose tracking the performance on the  two objectives of the reward function. This way, the human operators can have an insight in the performance of the used policy.

\section{Conclusion and Further Work}

In this paper, we propose a pricing agent and an appropriate simulation environment, which can be used for training and evaluation of the agent. We address the challenges of safety, robustness and sample efficiency of the pricing agent which can increase the cost of deployment in a physical environment.
After implementing the proposed pricing agent and evaluation of the results, we propose further training of the customer agents from the simulation environment. They would keep their original reward function, but now they will be trained in an environment where the price signal is responsive to their actions. This could further improve their performance in terms of reducing the peak energy demand. These additionally trained customer agents could amount to an environment for  evaluation of the pricing agent when the customers are fully responsive to the price signals. 

\small

\bibliography{references}
\bibliographystyle{iclr2020_conference}

\end{document}